\documentclass[namedreferences]{solarphysics}
\usepackage[hyperref,optionalrh,solaromanenum]{spr-sola-addons} 
\usepackage{graphicx}                    
\usepackage{color}                       
\usepackage{breakurl}                         

\newcommand{\oql}{O$^{7+}$/O$^{6+}$}
\newcommand{\velunit}{km~s$^{-1}$}

\begin{document}

\begin{article}

\begin{opening}

\title{Coronal sources and in situ properties of the solar winds sampled by ACE during 1999-2008}

\author{Hui~\surname{Fu}$^{1}$\sep
        Bo~\surname{Li}$^{1}$\sep
        Xing~\surname{Li}$^{2}$\sep\
        Zhenghua~\surname{Huang}$^{1}$\sep
        Chaozhou~\surname{Mou}$^{1}$\sep
        Fangran~\surname{Jiao}$^{1}$\sep
        Lidong~\surname{Xia}$^{1*}$
       }
\runningauthor{Fu et al.} \runningtitle{ACE solar wind sources}

   \institute{$^{1}$ Shandong Provincial Key Laboratory of Optical Astronomy and Solar-Terrestrial Environment,
   Institute of Space Sciences, Shandong University, Weihai, 264209, China\\
              $^{2}$ Department of Physics, Aberystwyth University, Ceredigion, Wales, UK, SY23
              3BZ\\
$*$ correspondence addressed to xld@sdu.edu.cn
             }

\begin{abstract}
We identify the coronal sources of the solar winds sampled by the ACE spacecraft
    during 1999-2008, and examine the in situ solar wind properties
    as a function of wind sources.
The standard two-step mapping technique is adopted to establish the photospheric footpoints
    of the magnetic flux tubes along which the ACE winds flow.
The footpoints are then placed in the context of EIT 284~\AA\
{images and photospheric magnetograms}, allowing us to
    categorize the sources into {four} groups: coronal holes (CHs),
    active regions (ARs), {the quiet Sun (QS), and ``Undefined''}.
This practice also enables us to establish the response to solar activity
    of the fractions occupied by each kind of solar winds,
    and of their speeds and \oql\ ratios measured in situ.
{
We find that during the maximum phase, the majority of ACE winds originate from ARs.
During the declining phase, CHs and ARs are equally important contributors to the ACE
    solar winds.
The QS contribution increases with decreasing solar activity, and maximizes in the minimum phase
    when QS appear to be the primary supplier of the ACE winds.}
With decreasing activity, the winds from all sources tend to
    become cooler, as represented by
    the increasingly low \oql\ ratios.
On the other hand, during each activity phase, the AR winds
{tend to be the slowest and}
    associated with the highest
    \oql\ ratios, and the CH winds correspond to the other extreme, with the QS winds lying in between.
Applying the same analysis method to the slow winds only, here defined as the winds with speeds lower than 500 km s$^{-1}$,
    we find basically the same overall behavior, as far as the contributions of individual groups of sources
    are concerned.
{This statistical study indicates that QS regions are an important source of the solar wind during the minimum phase.}
\end{abstract}
\end{opening}

\section{Introduction}
     \label{sec_intro}

Identifying the source regions of the solar wind is important both as
    a fundamental issue in solar physics~\citep{2012SSRv..172..169A}
    and from the space environment perspective~\citep[e.g.,][and references therein]{2002JGRA..107.1154L}.
This practice dates back to the era when the solar
    wind was first measured~\citep[][see also \citeauthor{2013JAdR....4..215P}~\citeyear{2013JAdR....4..215P} for a historic overview]{1966sowi.conf...25S,1973SoPh...33..241N,1998JGR...10314587N}.
With the solar wind data accumulated
    throughout several solar
    activity cycles
    in both near-ecliptic and polar orbits,
    scenarios have emerged as to how the solar wind sources evolve with solar activity.
This concerns not only the solar winds sampled by individual
    spacecraft but also the solar winds
    throughout the heliosphere \citep{2002JGRA..107.1154L}.

Traditionally, the studies on the solar wind sources start with categorizing
    the winds into the fast (with proton speeds $v$ over, say, 500~km\,s$^{-1}$)
    and slow ones ($v \lesssim 500$~km\,s$^{-1}$)
    \citep[e.g.,][]{2006SSRv..124...51S}.
Regarding the fast solar wind (FSW), the coronal source is generally accepted
    to be coronal holes~\citep[e.g.,][]{1973SoPh...29..505K,1977RvGSP..15..257Z,1999SSRv...89...21G}.
Tracing the wind sampled by \textit{Pioneer VI} and \textit{Vela},
\citet{1973SoPh...29..505K} were the first to associate the FSW
    with a coronal hole.
Then \citet{1977RvGSP..15..257Z} suggested that all coronal holes
    are sources of the FSW.
Using the SOHO/SUMER data, the outflows at the base of polar~\citep{1999Sci...283..810H}
    and equatorial~\citep{2003A&A...399L...5X} coronal holes were measured,
    with the results supporting the notion that the FSW originates in
    coronal funnels~\citep{2005Sci...308..519T}.
On the other hand, while examining the ACE and \textit{Ulysses} data
    for four Carrington rotations during the Cycle 23 maximum,
    \inlinecite{2002JGRA..107.1488N} concluded that a fraction of the FSW
    originate also from active regions.

The sources of the slow solar wind (SSW) are substantially more complex.
While there exists the consensus that the SSWs are associated with coronal streamers,
    debates remain as to exactly where in or around streamers the SSWs originate.
The scenario proposed for solar minimum conditions
    by~\inlinecite{1998ApJ...498L.165W}
    suggests that there are two kinds of SSWs, with one originating from streamer stalks
    and the other from just inside coronal holes and immediately adjacent to streamers.
The former source is consistent with the outmoving plasmoids found
by SOHO/LASCO \citep{1997ApJ...484..472S,1998ApJ...498L.165W},
    while the latter source is corroborated by the SOHO/UVCS
    measurements \citep{2010AdSpR..46.1400A}
    and consistent with the established inverse correlation of the flow tube expansion
    with the solar wind speed~\citep{1990ApJ...355..726W}.
However, even at solar minimum, this scenario remains to be
    complemented with the expected source
    of the SSWs from inside streamers, either via direct flow of the plasma from
    the magnetically open fields in streamer cores \citep{1997ESASP.404...75N}
    or via the evaporation of plasmas from the magnetic arcades in streamer
    helmets (\citeauthor{1999SSRv...87..323S},~\citeyear{1999SSRv...87..323S}, also~\citeauthor{2005JGRA..11012112L},~\citeyear{{2005JGRA..11012112L}}).
Besides, using the method of interplanetary scintillation (IPS) tomographic analysis,
    \citet{1999JGR...10416993K} found that yet another SSW source is the unipolar regions
    in the vicinity of active regions (ARs).
A further and more direct study associating the SSW with ARs comes with Hinode X-ray and EUV spectral observations,
    where the edge of ARs was shown to host persistent upflows
    with speeds reaching 100 km s$^{-1}$ \citep{2008ApJ...676L.147H}, which may
    account for up to 1/4 of the in situ SSW \citep{2007Sci...318.1585S}
    provided that these upflows eventually turn into outflows.
Indeed, \citet{2012SoPh..281..237V}~(also see~\citeauthor{2014SoPh..289.3799C},~\citeyear{2014SoPh..289.3799C}
    and \citeauthor{2014SoPh..289.4151M},~\citeyear{2014SoPh..289.4151M})
    showed that these upflows may access coronal magnetic fields that open into
    interplanetary space.
In addition, using X-ray high temporal-spatial resolution images,
    \citet{2010A&A...516A..50S} found that the magnetic
    reconnection of co-spatial open and closed magnetic field lines at
    coronal hole boundaries creates the necessary conditions for
    plasmas to flow to large distances.
This provides an explanation for largely-blue-shifted events observed
    with EIS/Hinode \citep{2012A&A...545A..67M}, indicating these plasma outflows
    are also a possible SSW source.
Comparing the remote sensing and in situ measurements,
    \citet{2005JGRA..110.7109F} suggested that the SSW may also arise from the quiet Sun.
When it comes to solar maximum conditions, SSWs are found to
    originate from small coronal holes and active regions where open
    magnetic field lines exist
    \citep{2002JGRA..107.1488N,2003ApJ...587..818W,2004SoPh..223..209L,2006ApJ...646.1275K,
    2006SSRv..124...51S,2009ApJ...691..760W}

While the identified coronal sources of the solar wind are diverse,
    there seem to be an agreement on the approaches behind the identification procedure.
First, unlike the solar wind speed itself, ionic charge states, especially those of
    oxygen and carbon, are suggested to be a telltale signature of the wind sources.
Take oxygen for example.
The abundance ratio \oql\ measured in the in situ solar wind is generally accepted to
    reflect the electron temperature in the coronal sources, given that it
    does not vary with distance beyond a fraction of a solar radius above the solar surface
    \citep{1983ApJ...275..354O,1986SoPh..103..347B,2000JGR...10510527H,2012ApJ...750..159L}.
Now that the temperatures are different in different coronal regions, a comparison of the
    in situ charge states then allows one to associate the in situ wind with a particular
    coronal source \citep[e.g.,][]{2000JGR...10518327Z,2001IAUS..203..585Z,2012ApJ...744..100L}.
With this spirit, \citet{2009GeoRL..3614104Z} divided the
    non-transient solar winds into two categories:
    those from coronal holes (CH winds) and those from outside coronal holes (non-CH winds)
    with \oql\ values lower and higher than 0.145, respectively.
As a result, about 42\% of the ecliptic solar wind was found to be of non-CH origin
    during 1998-2008.
Second, a model of coronal magnetic field is often indispensable.
For this purpose, while sophisticated Magnetohydrodynamic (MHD)
models are sometimes adopted \citep{2010AdSpR..46.1400A},
    the potential-field-source-surface model (PFSS) and its variants have been in much wider use.
On the one hand, this practice established the long-term trend of the wind speed being inversely
    correlated with the lateral expansion of the flow tubes \citep{1990ApJ...355..726W}.
On the other hand, applying the PFSS model with an archive of the
    synoptic magnetogram data
    leads \citet{2002JGRA..107.1154L} to the distribution of
    sources of the heliospheric solar wind as a function of solar activity
    for nearly three activity cycles.
In particular, \citet{2002JGRA..107.1154L} found that although
    polar coronal holes exist for more than 80\% of a solar cycle,
    they contribute to the ecliptic solar winds significantly
    only during half of a cycle.
During the other half of a cycle, the near-ecliptic winds originate from
    mid- and low-latitude sources instead.

Given the diversity of the wind sources and the complexity of the activity-dependence of
    these sources during a solar activity cycle,
    the present study is intended to examine, in a statistical manner,
    the fractions taken up by the in situ solar winds from various sources from the activity maximum to minimum in Cycle 23.
To this end, we start with the \textit{in situ} wind speed measurements, and adopt
    the standard two-step mapping procedure \citep{1998JGR...10314587N,2002JGRA..107.1488N,2004SoPh..223..209L}
    to trace the winds to their footpoints at the solar surface.
We then examine the corresponding coronal images recorded by SOHO/EIT
    {as well as photospheric magnetograms},
    and ask the question where the footpoints are located: are they located in a coronal hole (CH),
    an active region (AR), or the quiet Sun (QS)?
The solar winds are therefore grouped {accordingly},
    enabling us to address the question how their in situ properties
    differ and evolve with different activity levels.

Our study differs from previous studies with similar objectives or similar approaches in the following aspects.
First, the approach combining a footpoint tracing method with the context of coronal images
    follows closely the one in \citet{2004SoPh..223..209L}, which is in turn built on
    \citet{1998JGR...10314587N,2002JGRA..107.1488N} where the imaging data were not used.
However, while \citet{1998JGR...10314587N} focused on the Cycle 22--23 minimum,
    and \citeauthor{2004SoPh..223..209L} (\citeyear{2004SoPh..223..209L}, also \citeauthor{2002JGRA..107.1488N}~\citeyear{2002JGRA..107.1488N})
    concerned the Cycle 23 maximum,
    we examine the solar wind dataset that spans the interval from 1999 to 2008 in which
    the declining phase and Cycle 23--24 minimum are included.
Furthermore, the solar winds from sources other than CHs and ARs are
    paid attention to, and are classified as the QS winds.
Second, both this study and the one by \citet{2009GeoRL..3614104Z}
    (ZZF09 hereafter) have similar objectives in examining the
    distribution of wind sources in response to solar activity.
However, the approach for identifying the sources in this study is
    different from the one by ZZF09 where the \oql\ values are a primary discriminator.
We note that, given the uncertainties in both approaches, the results of this study
    are meant not to be contrasted with but rather to complement ZZF09, with the hope that
    new light can be shed on the sources of the near-ecliptic solar winds.
Third, while both using the PFSS model and being statistical in nature,
    our study differs from the one by \citet{2002JGRA..107.1154L}
    in that we also employ the
    {imaging as well as magnetogram data to} classify the sources
    instead of {using} the locations relative to the equator
    as in \citet{2002JGRA..107.1154L}.
Fourth, given the considerable interest in and the complexities
    associated with the sources of the slow solar wind,
    we will analyze the ACE solar winds in general, and examine the slow ones in particular.
In Section~\ref{sec_data}, we describe the data and our method of
analysis. The results are then given in Section~\ref{sec_results}.
Section~\ref{sec_conclusion} summarizes the present study, ending
with some concluding remarks.

\section{Data and analysis} 
      \label{sec_data}
The two-step mapping procedure used in the present study closely follows the one in
    \citet{1998JGR...10314587N,2002JGRA..107.1488N,2004SoPh..223..209L}.
To initiate the procedure,
    we use daily averages of the solar wind speed made with
    the \textit{Solar Wind Electron, Proton, and Alpha
    Monitor} (SWEPAM, \citeauthor{1998SSRv...86..563M},
    \citeyear{1998SSRv...86..563M}) on board
    the \textit{Advanced Composition Explorer} (ACE, \citeauthor{1998SSRv...86....1S}, \citeyear{1998SSRv...86....1S}).
Also used are the daily averages of
    the abundance ratios \oql\ recorded by
    the \textit{Solar Wind Ion Composition Spectrometer}
    (SWICS, \citeauthor{1998SSRv...86..497G}, \citeyear{1998SSRv...86..497G}),
    and the magnetic field measurements with the \textit{Magnetic Field Experiment}
    (MAG, \citeauthor{1998SSRv...86..613S} ,\citeyear{1998SSRv...86..613S}).
Given that we are interested in the non-transient solar winds,
    one immediate purpose for using the \oql\ ratios
    is to eliminate from the ACE dataset those intervals occupied by
    interplanetary \textit{Coronal Mass Ejections} (ICMEs).
To do this, we adopt the same approach as
    in~ZZF09~\citep[see also][]{2004JGRA..109.9104R} whereby
    we discard the data with \oql\ ratios exceeding
    $6.008 \exp(-0.00578 v)$, in which $v$ is the wind speed in \velunit.
A detailed analysis by ZZF09 shows that
    this criterion adequately separates ICMEs from the non-transient ambient winds,
    being reliable in 83.2\% of the cases examined therein.
The \textit{in situ} data used in this study span the years between 1999 and 2008,
    hence encompassing nearly half of the Cycle 23.

The mapping procedure involves two steps.
First, the loci of the solar winds are found on the source surface, placed at a heliospheric distance
    of 2.5~$R_\odot$ as implemented by the coronal magnetic field model.
This is done via a ballistic approach,
    whereby the longitude correction due to solar rotation is determined by
    the time for a wind parcel to travel from the source surface
    to the spacecraft.
Here a constant wind speed is used, and assumed to be the one measured by ACE/SWEPAM.
The wind parcel is then traced from the source surface to the photosphere
    by following the magnetic field lines computed by using a PFSS model, provided in the
    PFSS package as part of the Solar Software. Instead of using the
    synoptic magnetograms as was done in e.g., \citet{1998JGR...10314587N},
    this package uses, as the boundary data,
    the magnetograms measured with SOHO/MDI which are updated every 6 hours.
It outputs the magnetic field vector on a $39 \times 384 \times 192$ grid
    in spherical coordinates inside the source surface
    (for details, see \citeauthor{2003SoPh..212..165S}, \citeyear{2003SoPh..212..165S}).
It should be noted that as implied by the mapping procedure,
    the magnetic polarity at the field line footpoint needs to be checked against the
    one measured in situ.
\citet{2003SoPh..212..165S} found that during 1997-2001, 83\% of
    footpoint polarities matched
    the interplanetary magnetic field (IMF) measurements at the Earth.
In this work, we find a similar behavior: the footpoint polarities
    are consistent with what is measured by ACE/MAG in 81\% of the data
    from 1999 to 2008.
To ensure consistency, we do not include in our further analysis those dates when
    the polarities at the two ends of the mapping procedure do not match.
Table~\ref{tbl_polarity_all} presents, as a function of time, the number of daily samples of
    the non-transient solar wind
    (second column, labeled ``All sources''), which is sub-divided into the counts
    of the solar winds from
    {CHs (third column), ARs (fourth), QS (fifth), and Undefined sources (sixth).}
Given in the parentheses are the numbers that correspond to the
cases where the magnetic polarities match. In total, during
1999-2008, {2124} samples are examined in our further analysis,
    among which {615 (803, 425)} samples are associated with CHs (ARs, the QS).
One can see that a significant mismatch takes place in 2007 and
2008. This is possibly understandable given that, close to the
cycle minimum,
    the ACE spacecraft was close to the heliospheric current sheet.
When mapping the winds to the source surface, a small uncertainty
may lead to a wrong polarity. For future reference,
Table~\ref{tbl_polarity_slow} presents the comparison between the
footpoint polarity
    and the in situ one for the slow solar winds with speeds lower than 500~\velunit.

The footpoints are then placed in the context of photospheric
magnetograms and the EUV images taken by
    the \textit{Extreme ultraviolet Imaging Telescope} (EIT, \citeauthor{1995SoPh..162..291D},
    \citeyear{1995SoPh..162..291D}) onboard
    SOHO (\textit{Solar and Heliospheric Observatory}, \citeauthor{1995SoPh..162....1D}, \citeyear{1995SoPh..162....1D}).
While EIT operates at a number of passbands
    (Fe~{\sc{ix/x}} 171~\AA, Fe~{\sc{xii}} 195~\AA, Fe~{\sc{xv}} 284~\AA\ and He~{\sc{ii}} 304~\AA),
    we choose the 284~\AA\ one because the images recorded in this passband
    reflect the corona at the highest altitude such that
    coronal holes are more visible.
In this passband, EIT takes full-Sun images with a pixel size of
2.6$''$
    four times a day.
For consistency, the field line footpoints are compared with images taken at around 13:00\,UT on
    the day corrected for the wind travel time.

\begin{figure}    
   \centerline{\includegraphics[width=0.8\textwidth,clip=]{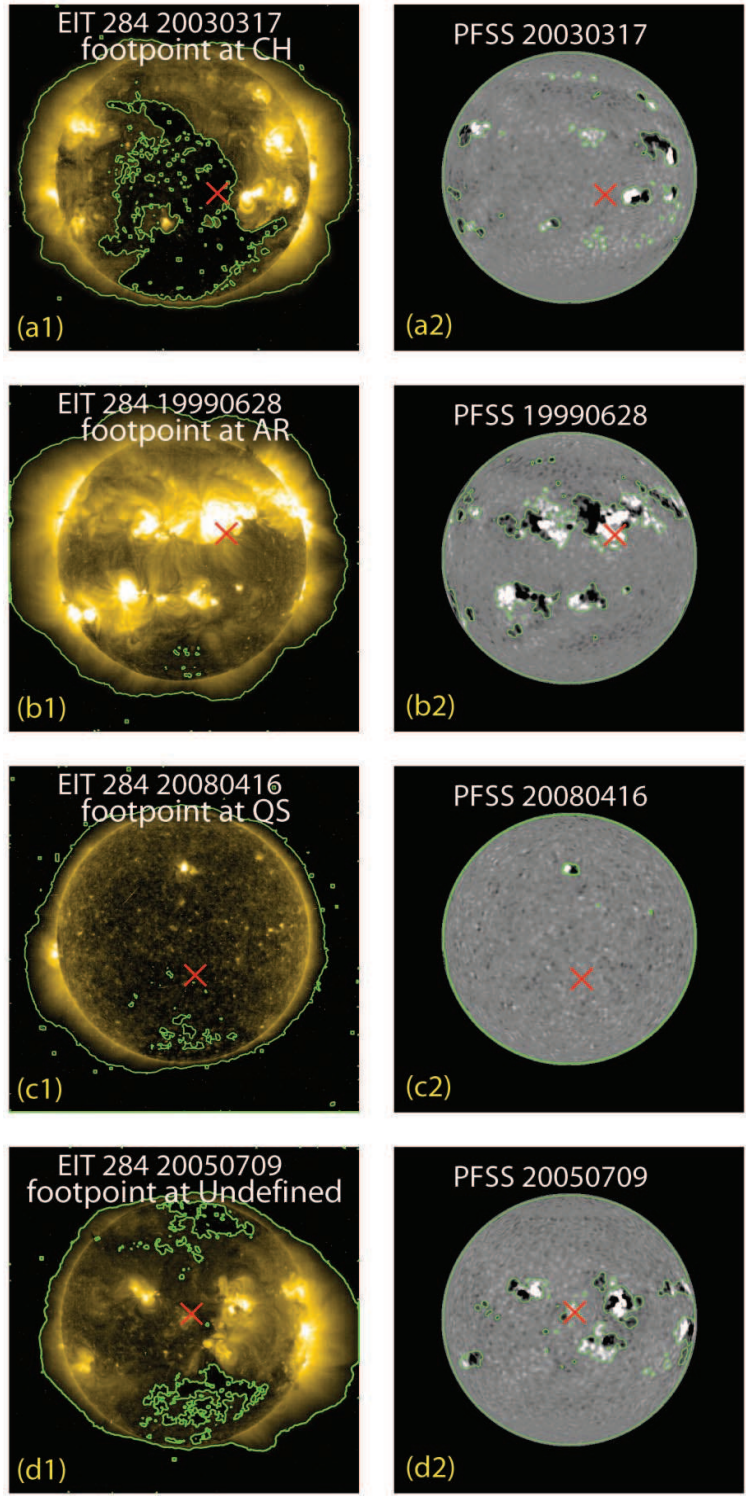}
              }
              \caption{{Illustration of the classification scheme of the ACE solar winds.
              The footpoints of the solar wind flow tubes are given by the red crosses, which are classified as being associated with
                  a coronal hole (the first row), an active region (second), the quiet Sun (third) {and some undefined source (bottom).
              The left column presents the EIT 284~\AA\ images, while the right column
                  gives the corresponding magnetic morphology of the photosphere.
              The green contours outline CH (left column) and Magnetically Concentrated Area (MCA) boundaries (right)}.
              An animation showing the sources during 1999--2008 is available online.
                      }}
   \label{fig_src_id}
   \end{figure}

The classification scheme is illustrated in Figure~\ref{fig_src_id},
    where the EIT images (the left column) are overplotted with
    the footpoint locations represented by the red crosses.
The photospheric magnetograms, on which our scheme also relies,
    are derived from the PFSS model and given in the right column.
The scheme is detailed as follows.

\begin{figure}    
   \centerline{\includegraphics[width=1.0\textwidth,clip=]{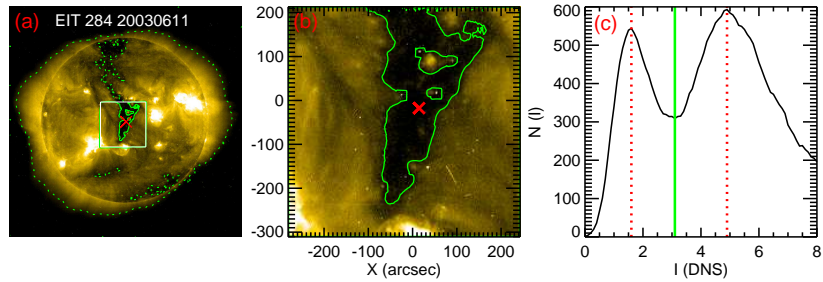}
              }
              \caption{The scheme for outlining coronal holes.
              Panel (a) presents the EIT 284 image on 2003 June 11, when the traced-back footpoint (the red cross)
                 is located close to a low-latitude CH.
              The white box encloses the region for which the intensity histogram is constructed
                 and presented in panel (c), where the solid green line represents the minimum
                 between the two peaks, given by the two red dotted lines.
              This minimum is used as the threshold to delineate CH boundaries in (a),
                 where the contours inside (outside) the box are given by the solid (dotted) lines.
	      Panel (b) is an enlarged version of the part enclosed by the box in (a).
               }
   \label{fig_CH_histo}
   \end{figure}

A quantitative approach for identifying coronal hole boundaries is implemented,
    and the winds that have footpoints located within the hence identified coronal holes (CHs) are classified as ``CH
    winds'' accordingly.
This approach, which largely follows that in \citet{2009SoPh..256...87K} and \citet{2014ApJ...787..121K},
    is illustrated in Figure~\ref{fig_CH_histo}.
If a footpoint is located inside or close to an apparently dark area, then a rectangular box
    (the white box in Figure~\ref{fig_CH_histo}a)
    is chosen to enclose this part of the dark region and its surrounding area.
An intensity histogram is constructed, and a multi-peak distribution is then
    obvious (Figure~\ref{fig_CH_histo}c).
The well-defined minimum between the first two peaks then defines the threshold for identifying the
    CH boundary (see the contours in Figure~\ref{fig_CH_histo}a, also Figure~\ref{fig_CH_histo}b, which is the enlarged version of the part inside the box).
On the one hand, this scheme enables one to objectively define CH boundaries using the EUV images in only one passband;
    on the other hand, it is not influenced by the variation of coronal emissions with solar activity.

A description of some technical details for implementing this scheme seems necessary.
In practice, we started with asking the question whether there is a dark region close to the traced-back footpoint.
By ``close'', we mean roughly ``within 100 arcsecs''.
If the answer is Yes, we then draw a rectangular box, varying in size but typically a few hundred arcsecs across,
    which encloses both a substantial part of the dark region and its surrounding area.
The footpoint is always within this box.
It turns out as long as the box is sufficiently large, its size does not significantly influence
    what one identifies as CH boundaries, for the minima in the different histograms
    pertinent to different box sizes do not differ substantially.
If the answer is NO, we visually choose the dark area that is the closest to the footpoint,
    and use the same approach to delineate the CH boundary (see e.g., Figure~\ref{fig_src_id}d1).
If there is no large apparent EUV CH altogether, then we use the threshold found for some obvious
    CH one or a few days prior to this particular day
(An example is shown in Figure~\ref{fig_src_id}b1, where the CHs near the two poles are contaminated so significantly that
    the minimum between the first two peaks in the intensity histogram can hardly be discerned).
The box is substantially smaller than the disc size, we nonetheless use the threshold to delineate
    CH boundaries throughout the entire solar disc.
A space-dependent threshold may be more accurate for mapping CH boundaries on the entire disc,
    but our approach suffices given that our purpose is to examine whether the footpoint is located inside a CH.
Besides, as illustrated by Figure~\ref{fig_CH_histo}a, while a single threshold is adopted, the contours outside the box
    (the dotted lines) also outline CHs rather accurately.

Our association of a footpoint with an Active Region (AR) or the quiet Sun (QS) relies on the magnetic morphology
    of the photospheric regions embodying the identified footpoints.
The most obvious features on the photosphere are magnetic clusters,
which are tentatively named ``magnetically concentrated area
    (MCA)''.
Intuitively speaking, MCAs correspond to strong magnetic fields.
To make this definition more objective, the absolute value of the radial component of the
    photospheric magnetic field $|B_{r, \odot}|$ computed from the PFSS model is used.
We experimented with different contour levels, $|B_{r, \odot}|_{B}$, used for outlining MCAs
    presented in Figure~\ref{fig_src_id} and the attached movie.
In practice, if $|B_{r, \odot}|_{B}$ is assigned a value $1.5-4$
times the mean of $|B_{r, \odot}|$, then MCAs become well defined.
That the MCA morphology is not sensitive to some given $|B_{r,
\odot}|_{B}$, as long as it is in the mentioned range, suggests that
MCAs have sharp boundaries. This is understandable considering that
MCAs have a strong spatial gradient in $|B_{r,\odot}|$. The
thus-defined MCAs encompass all the active regions numbered by NOAA
as provided by solarmonitor \footnote{http://solarmonitor.org}.
However, not all MCA patches correspond to a numbered AR. Many of
these turn out to correspond to plages with magnetic field weaker
than concurrent numbered ARs (see section 3.2
in~\citeauthor{2005SoPh..228..361Z}~\citeyear{2005SoPh..228..361Z}).
In the three solar activity phases (solar maximum, decline, minimum,
see Figure~\ref{fig_perc_all}) the contour levels are
    chosen to be 2.0, 2.5, and 3.0 times the mean of $|B_{r,\odot}|$, respectively.
In the MAX and DEC phases the threshold is about 10-20~Gs, which is
close to the
    lower bound (15 Gs) adopted by~\citet{2009ApJ...691..760W} to identify slow solar winds from ARs.

With CH boundaries quantitatively defined, when identifying AR and QS sources we need only to concern
    about the regions outside CHs.
An AR source is defined when a footpoint is located inside an MCA
    that is a numbered AR by NOAA.
Likewise, QS sources are defined when a footpoint is located outside any MCA.

With the present grouping scheme, what is unclassified is then named ``Undefined'',
   and corresponds to the case where a footpoint is inside some MCA that is not numbered by NOAA.
These sources may be associated with a decaying/developing AR,
   but it is also possible that they are distinct from AR sources (see, e.g., 06/25-06/27 2005
   in the movie where the source is likely a QS one). This is why the word ``Undefined'' is chosen.

Such a scheme will not overestimate the counts in the respective groups.
First, the counts of AR and QS sources are not overestimated, since some footpoints
    deemed ``Undefined'' may in fact be AR and QS sources.
Second, the counts of CH winds are not overestimated either, for the
current definition of CHs excludes a fraction of CHs with overlying
bright emissions. In any case, the counts in the Undefined group
account for only a minor fraction of the samples
   (11.2\%, 9.5\%, 14.5\%, 10.4\%, 9.3\%, 18.5\%, 17.9\%, 19.5\%, 18.0\%, and 5.3\%
   for the years 1999 to 2008, respectively).

Before proceeding, several remarks on our approach seem in order.
The first remark is on the reliability of the PFSS model, given its apparently oversimplification
    of imposing a spherical source surface, neglecting volumetric electric currents
    between the source surface and the photosphere, and supposing purely radially directed field lines
    outside the source surface.
Nonetheless, a detailed comparison study by
    \citet{2006ApJ...653.1510R} demonstrated that the magnetic field configuration
    computed by the simple PFSS model agrees well with the one found in sophisticated MHD computations,
    provided that both models are driven by the same line-of-sight magnetograms.
From the practical point of view, the magnetically open regions obtained by the PFSS model
    well match the coronal hole regions identified in, say,
    the He I 10830 synotpic diagrams
    \citep{1982SoPh...79..203L,1998JGR...10314587N,2002JGRA..107.1488N,2003SoPh..212..165S}.
A good way to make sure that the traced-back footpoints are reasonably accurate is to
    compare the current SolarSoft PFSS results with some other calculations.
To address this, we randomly chose three Carrington Rotations in the
MAX, DEC, and MIN phases, and
    compared our derived footpoints with those derived from the PFSS model where the magnetogram input is from the Wilcox Solar Observatory (WSO).
We found that the fraction of the days when the two different sets
of footpoints belong to the same
    open field region is 80.7\% for CR 1969, 84\% for CR 2005, and 82.6\% for CR 2054.
Nevertheless, let us stress that the fraction that the two sets do agree is substantial enough that
    the statistical study we conduct can be deemed reliable.

Another source of uncertainty may come from the mapping procedure, particularly in view of the simple ballistic
    treatment involved in the first step.
As demonstrated by \citet{1973SoPh...33..241N} (also see
     \citeauthor{2002JGRA..107.1488N},~\citeyear{2002JGRA..107.1488N}),
while the solar wind may experience
    some acceleration beyond
    the source surface, this effect may be counter-balanced by the near-Sun corotation.
Actually a further evidence lending us confidence with this mapping procedure is that,
    when inspecting the footpoints on a consecutive basis (please see online animation 1 attached to Figure~\ref{fig_src_id}), one can see an orderly distribution
    of the locations of footpoints.
They stay in a particular group for several days before moving to another group.
In addition, assuming that the uncertainty with the source longitude at the source surface
    is $\pm 10^\circ$ \citep{1973SoPh...33..241N}, we select two Carrington rotations in each sub-interval
    (see Figure~\ref{fig_perc_all}a) and examine how well our classification scheme works.
This is done by tracing the photospheric footpoint from a locus on
    the source surface $10^\circ$ eastward or westward of the nominal locus,
    and then examining whether the footpoint is located in a different area in the EIT images.
We found that at maximum activity, about 30\% of the footpoints indeed are associated with an area different from
    what we identified using the nominal locus.
During the declining and minimum phases, however, this mismatch reduces to $\lesssim 20\%$ and $10\%$ of the cases examined,
    respectively.

\section{Results}
 \label{sec_results}

\subsection{Sources of the ACE wind between 1999 and 2008}
 \label{All_range}
\begin{figure}    
   \centerline{\includegraphics[width=1.0\textwidth,clip=]{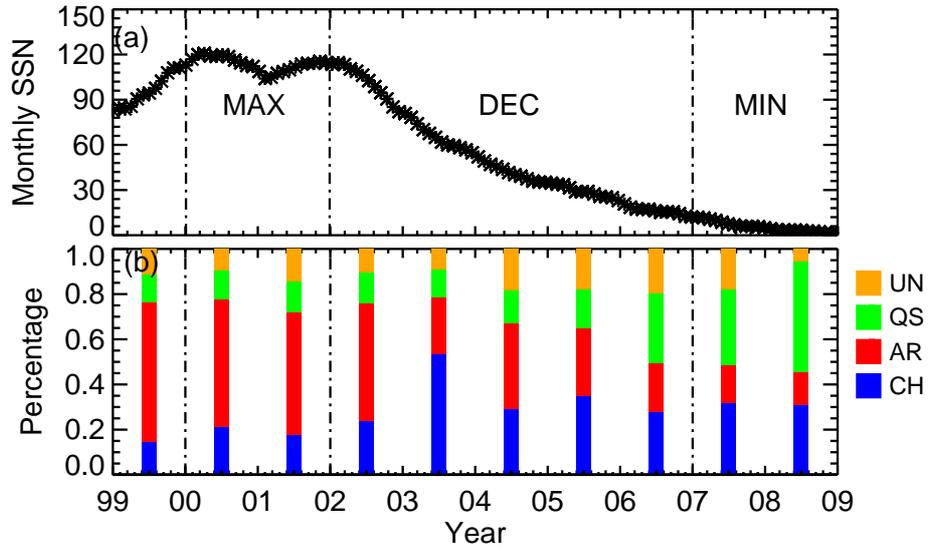}
              }
              \caption{Fractions of the ACE winds with different sources as a function of time.
              Panel (a) shows the temporal evolution of the smoothed monthly sunspot number during 1999-2008,
              which is further divided into the maximum (labeled MAX), declining (DEC)
              and minimum (MIN) phases.
              Panel (b) gives the percentage of the coronal hole (CH, blue),
                 active region (AR, red), quiet Sun (QS, green) {and undefined (UN, orange)} winds.
               }
   \label{fig_perc_all}
   \end{figure}

Having categorized the winds, we then address the question of the percentage each kind of wind
    occupies, and how this evolves in response to solar activity.
Figure~\ref{fig_perc_all}a presents the monthly average of the smoothed
    sunspot numbers from 1999 to 2008.
Three sub-intervals, labeled MAX, DEC and MIN,
    are then defined according to the level of solar activity.
Figure~\ref{fig_perc_all}b presents the percentage of the
CH (shaded blue), AR (red), QS (green) and Undefined (orange) winds in this period.
These percentages are yearly averaged values, and add up to unity in each year.
One can see that in the sub-interval MAX, the QS supplies only a small fraction of the winds sampled
    by ACE ($\sim 10$\%), the contributions from CHs is $\sim 15-20$\%,
    and more than half (56\%) of the ACE winds originates from ARs.
In the declining phase (2002 -- 2006), the contributions from CHs and ARs both amount to roughly
    34\%, and the contributions from ARs (QS) tend to decrease (increase).
As for the sub-interval MIN (2007 -- 2008), the fraction of the
    winds from ARs is only marginal ($\lesssim 17\%$),
    while some 31\% comes from the CHs, and nearly half of the winds
    originates from the QS.

Despite the differences in the approaches for identifying the solar wind sources,
    Figure~\ref{fig_perc_all} agrees with Figure~1 in ZZF09 in that
    there exists a tendency for coronal sources other than CHs
    to contribute significantly to the ACE solar winds between 1999 and 2008.
Overall, Figure~\ref{fig_perc_all}b indicates that non-CH sources may be more important than CHs
    in terms of their mass supply to the solar wind.
In particular, Figure~\ref{fig_perc_all}b indicates that the majority of the near-Earth solar wind
    comes from ARs during the Cycle 23 maximum.
This behavior agrees with ZZF09, and seems to persist to the Cycle 24 maximum as indicated by
    the very recent study by~\citet{2015NatCo...6E5947B}.
Furthermore, Figure~\ref{fig_perc_all} indicates that CH winds tend to dominate in the
    year of 2003, which is also in line with ZZF09.
What is new in Figure~\ref{fig_perc_all} is that
    {combining the imaging as well as the magnetogram data
    allows} us to further determine, among the sources outside CHs,
    the fractions of the winds from ARs and the QS.
Some apparent differences from ZZF09 arise as a result.
Around the Cycle 23--24 minimum, while ZZF09 indicated that CH winds dominate,
    our Figure~\ref{fig_perc_all}b suggests that the contribution from the QS is more important.
We stress that this apparent discrepancy stems from the differences in the schemes
    for classifying the solar winds.

\begin{figure}    
   \centerline{\includegraphics[width=1.0\textwidth,clip=]{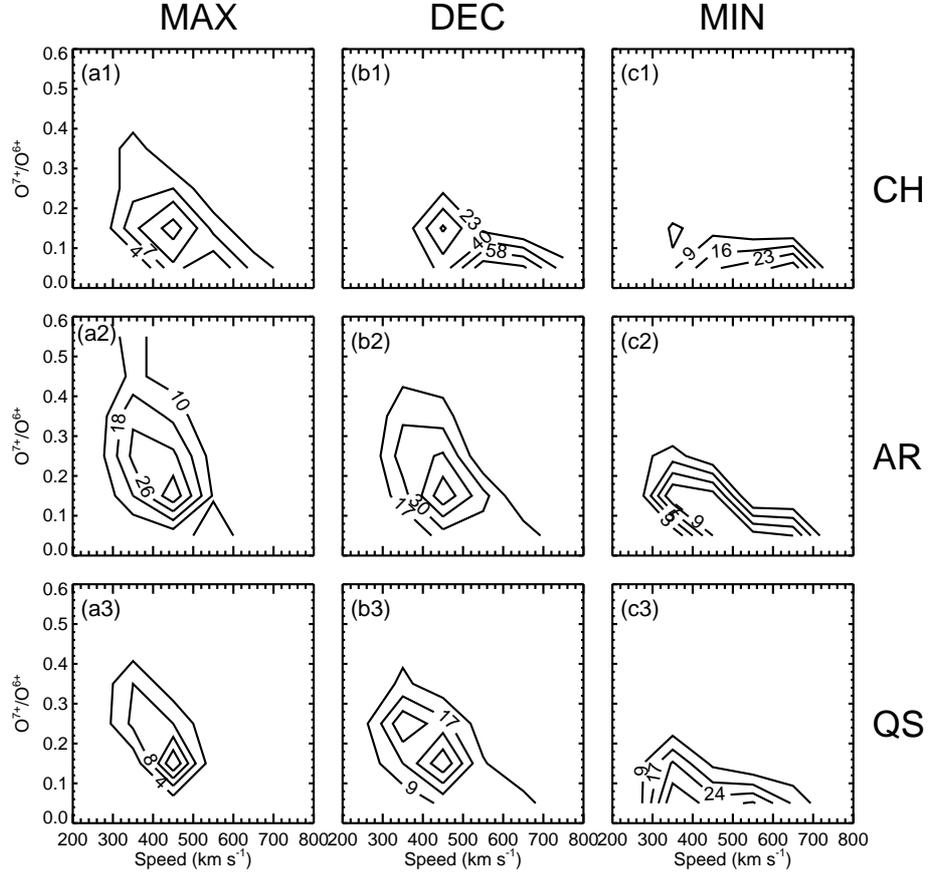}
              }
              \caption{Dependence on solar activity of the distribution of solar winds from different sources
              in the speed--\oql\ space.
              The left (middle, right) column corresponds to the maximum (declining, minimum) phase,
                 while the first (second, third) row represents the winds from coronal holes (active regions, the quiet Sun).
              Here the counts of solar wind samples in different groups are shown as contour plots with the contours equally spaced
                 in each panel.
              }
   \label{fig_contour}
   \end{figure}

More insights can be gained by examining how the in situ properties of the solar winds
    categorized by our scheme depend on solar activity.
We sort the wind speeds $v$ into $6$ bins uniformly
   spaced between 200 and 800 \velunit,
   group the \oql\ ratios into $6$ bins uniformly spaced between 0.0 and 0.6,
   then present in Figure~\ref{fig_contour} a contour plot
   in the $v$-\oql\ space the counts of the winds
   from different sources as labeled.
The left, middle, and right columns correspond to the intervals MAX, DEC, and MIN, respectively.
Consider the interval MAX (left column) first.
One notices that the majority of the winds corresponds to an \oql\ ratio larger than 0.145,
    which we recall is the criterion that ZZF09 employed to separate CH winds from non-CH ones.
However, a more detailed analysis like ours indicates that
    not all winds that have \oql\ ratios lower than the nominal value of $0.145$
    are from CHs.
Conversely, winds with \oql\ exceeding $0.145$ are not necessarily non-CH ones.
Now consider the years 2007 and 2008, labeled MIN.
One can see from Figure~\ref{fig_contour} (right column) that the
    \oql\ ratios tend to be low, with the majority
    being lower than 0.145, meaning that if categorizing the ACE winds by this threshold,
    one would find that nearly all the winds are from CHs.
However, combining the footpoint tracing approach with the EUV and magnetic field data,
    we find that the QS is the primary contributor to the ACE winds during this period.
Furthermore, comparing Figures~\ref{fig_contour}c1 with ~\ref{fig_contour}c3,
    one can see that the QS winds are distinct from the CH winds in that they tend to be substantially slower.
To select the proper subset of the fast solar wind sampled by ACE that comes from CHs,
    it would be almost unmistakable to choose those with speeds higher than 600~\velunit\
    and \oql\ lower than, say, $0.05$.
The contamination from the QS winds would be at most marginal, and
    that from the AR winds would be minimal.
We note in passing that this practice has been successfully
    employed by \citet{2014ApJ...781..110Z}.
Regarding the declining phase (middle column of
     Figure~\ref{fig_contour}),
     one finds that the possibility of distinguishing between CH winds
     and non-CH winds lies in between the extremes of maximum and minimum conditions.
This is particularly true in the speed dimension.
The CH winds tend to be faster than the non-CH ones (mainly from ARs in this case),
     and the difference between the two tends to be more obvious than
     for the MAX phase, but appears significantly less obvious than for the MIN phase.

The \oql\ ratios for the CH winds during the MAX phase (Figure~\ref{fig_contour}a1)
    require some explanation.
There appears to be a fraction of the CH winds for which the \oql\ values
    exceed 0.26.
If assuming ionization equilibrium, this would correspond to a
    freeze-in temperature exceeding 1.58~MK
    \citep{1998A&AS..133..403M}.
This is beyond the currently accepted electron temperatures
    derived from remote sensing measurements
    for CHs below 1.6~R$_\odot$ (\citealp{1993ApJ...413..435H,2000ApJ...532L..71E} and references therein).
This apparent discrepancy is not too worrisome given that
    this fraction of the CH winds tends to originate from the
    boundaries between CHs and ARs, while
    the measurements made by~\citet{1993ApJ...413..435H} pertain to the region
    well inside a polar CH.
Furthermore, as proposed by
    \citet{2000ApJ...532L..71E}, the electron
    distribution function may rapidly develop
    a non-Maxwellian character within the first several solar radii that eventually forms what
    is measured in situ as the halo electrons \citep{2006LRSP....3....1M}.
It is worth noting that this non-Maxwellian character is also possible to develop in
    AR and QS winds.

\begin{figure}    
   \centerline{\includegraphics[width=1.0\textwidth,clip=]{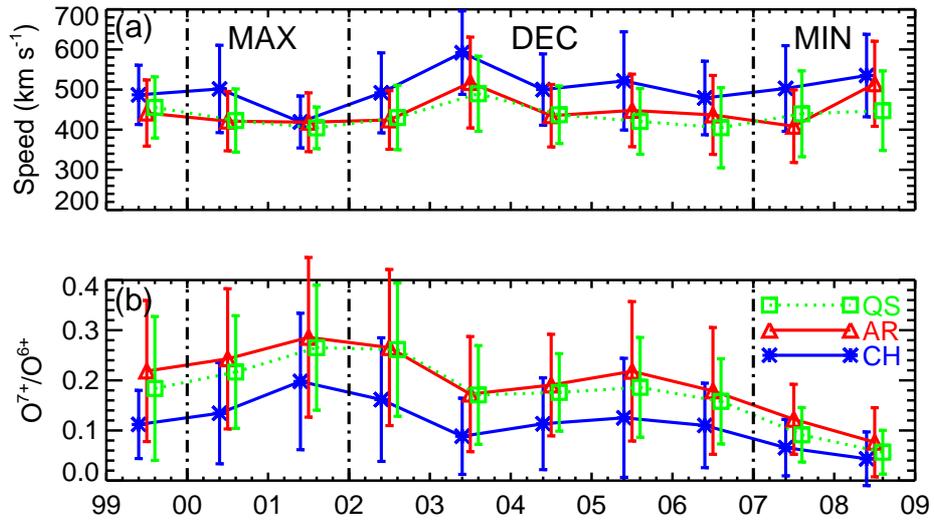}
              }
              \caption{In situ properties of solar winds from different sources as a function of time during 1999--2008.
              Here panels (a) and (b) are for the wind speeds and \oql\ ratios, respectively.
              The interval between 2000 and 2008 is further divided into three activity phases: maximum (MAX),
                  declining (DEC) and minimum (MIN).
              The winds from coronal holes (CHs), active regions (ARs) and the quiet Sun (QS) are represented by
                  the blue, red, and green curves, respectively.
              As for the error bars, they represent the standard deviations in each year.
              }
   \label{fig_tim_dep_all}
   \end{figure}

The differences in the in situ properties of the winds from different sources are further examined
    in Figure~\ref{fig_tim_dep_all},
    where (a) the wind speed  and (b) the oxygen charge state ratio are plotted as a function of time.
Given by the green, red, and blue curves are the parameters of the QS, AR, and CH winds, respectively.
The standard deviations are given by the error bars for the corresponding values,
    which are slightly displaced from one another
    for display purposes.
An immediate impression from Figure~\ref{fig_tim_dep_all} is that
    the CH winds tend to be the fastest, while the AR and QS winds have almost the equal speeds.
And the \oql\ ratios are lowest (largest) for CH (AR) winds, with
the QS winds lying in between.
However, the considerable overlap in either the speed or \oql\ ranges means that
    neither of the two parameters, on its own, seems to suffice to discriminate the wind sources.
Regarding the activity-dependence of the parameters,
    one can see from Figure~\ref{fig_tim_dep_all}a that the wind speeds in the three categories
    show a similar non-monotonic behavior.
Take the CH winds for instance.
Their speed start from relatively low values
{($\sim 500$~\velunit) around MAX, rise to some
     590~\velunit\ in 2003 before decreasing to around ${500}$~\velunit} in 2004-2006,
    and then gradually increase to some $550$~\velunit\ toward the MIN phase.
Moving on to Figure~\ref{fig_tim_dep_all}b, one can see that the
    overall tendency of \oql\ ratios in response to solar activity
    is opposite to that of the speed, as would be expected given the well-established
    inverse correlation of the two parameters \citep{2003ApJ...587..818W,2009ApJ...691..760W}.
Nonetheless, one can see that the \oql\ ratios from different
    groups of winds
    differ more significantly than the speeds do:
{Note the marked difference in the \oql\ values in the CH winds
    from those in the AR winds in the whole period.
It is noteworthy that with decreasing activity, the \oql\ ratios
    in all three types of solar winds tend to decrease,
    which agrees with \citet{2013ApJ...768...94L}.
The \oql\ values in AR (CH) winds are 0.26 (0.16) during MAX, and decrease to 0.10 (0.05) during MIN.
    }

\subsection{Sources of the ACE slow wind between 1999 and 2008}
 \label{SSW}

\begin{figure}    
   \centerline{\includegraphics[width=1.0\textwidth,clip=]{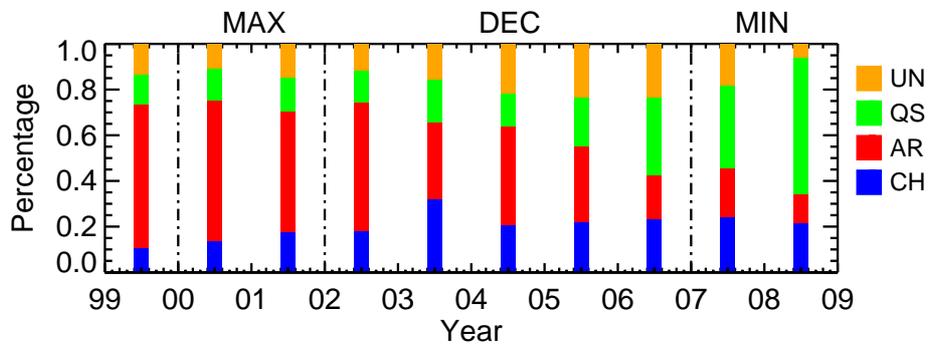}
              }
              \caption{Similar to Figure~\ref{fig_perc_all} but restricted to the slow wind with speeds less than 500\velunit.
              }
   \label{fig_perc_slow}
   \end{figure}

Given the considerable interest in understanding the origins of the slow solar wind (SSW),
    it is informative to apply the same practice to the slow winds alone.
In the present study, a SSW is defined to be the wind with speeds
    lower than 500~\velunit. Figure~\ref{fig_perc_slow} examines the
    time evolution of the fractions of the SSWs
    coming from various sources during 1999-2008.
Overall, the impression in the MAX and DEC phases is similar to what one finds in Figure~\ref{fig_perc_all}
    where the solar winds as a whole were considered.
This similarity to Figure~\ref{fig_perc_all}b is not surprising given that,
    as shown in Figure~\ref{fig_contour},
    most of the solar winds is on the lower side when the speed is concerned.
{During the MIN phase, the contribution from the QS to the slow wind
    is even more important than that to the overall solar wind.}
This is also understandable in view of Figure~\ref{fig_contour}c1, given that
    the solar winds from CHs are largely fast ones.

\begin{figure}    
   \centerline{\includegraphics[width=1.0\textwidth,clip=]{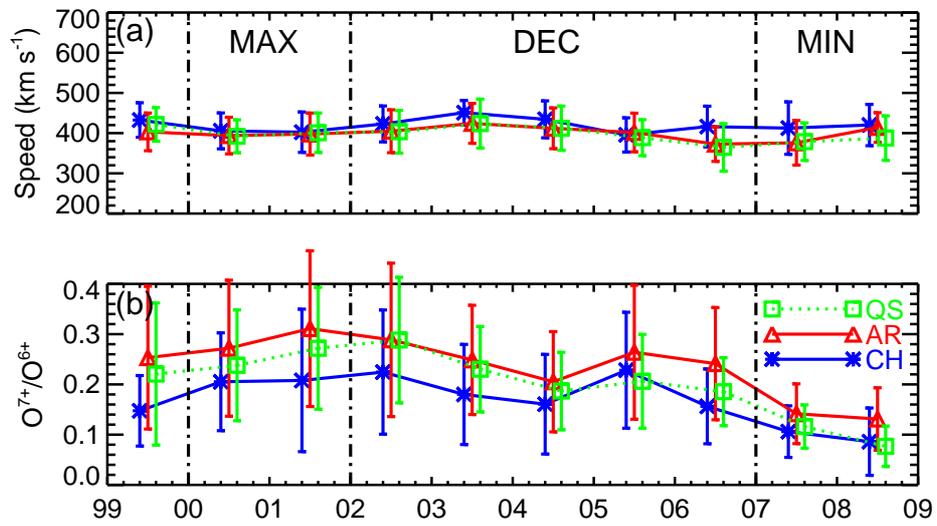}
              }
              \caption{Similar to Figure~\ref{fig_tim_dep_all} but restricted to the slow wind.
              }
   \label{fig_tim_dep_slow}
   \end{figure}

Figure~\ref{fig_tim_dep_slow} presents (a) the wind speeds and (b) the \oql\ ratios for the slow solar wind
    as a function of time.
As far as the wind speeds are concerned, one can see that the speed in a given group does not show a systematic variation with
    solar activity.
In addition, there is no clear-cut difference in the speeds of the winds
    from different groups.
A stronger temporal variation and a more significant difference in different groups of winds
    lie in the \oql\ values (Figure~\ref{fig_tim_dep_slow}b).
Overall, the \oql\ values for all the winds show a decrease with decreasing solar activity,
    and they are substantially different for different groups.
{The differences in the \oql\ values in winds from different sources
    may be a result of the intrinsic difference in the respective source properties,
    the magnetic field strength being the most likely one.
At any rate, this} reinforces the notion raised by \citet{2012SSRv..172..169A},
    who suggested that
    the compositional properties and temporal variability serve better
    in differentiating the wind sources than the speeds.

The SSW properties may be compared with previous studies.
\citet{2009ApJ...691..760W} suggested that the slow wind during
    1998-2007 mainly contains two components:
    one from small holes located in and around ARs with high \oql\ ratios during maximum,
    the other from the boundaries of large CHs with intermediate \oql\ values.
{
Our approach suggests that the majority of the former component indeed comes from ARs during maximum.
However, the latter component may actually come from all the three kinds of sources
    (see Figure~\ref{fig_contour}).}

\section{Conclusion}
\label{sec_conclusion}

The main purpose of this work is to examine, in a statistical sense, the sources
    of the solar wind sampled by ACE during 1999-2008 in general, and those of the slow solar wind in particular.
To this end, we start with the in situ wind speed, and find the photospheric footpoints of the wind parcels
    by employing the standard two-step mapping procedure \citep{1998JGR...10314587N,2002JGRA..107.1488N}
    where the Potential Field Source Surface (PFSS) model \citep{1969SoPh....6..442S,1969SoPh....9..131A} is used.
{We then associate the footpoints with various
    areas in the EUV images recorded by EIT in its 284 \AA\ passband and photospheric magnetograms.
With this association we} classify the ACE winds into three groups: coronal hole (CH), active region (AR), and quite Sun (QS)
    winds.
Our main results can be summarized as follows.

\begin{enumerate}
 \item
 {
 During Cycle 23 maximum (years 2000 and 2001), ARs are the main contributor to the ACE winds, the
    contribution of CHs (QS) is $\sim 20$\% (13\%).
 The winds in this interval tend to be slow, and the AR winds correspond to substantially higher \oql\ values than the CH winds.
 During the declining phase, the contributions from CHs and ARs both amount to roughly one third.
 Overall, the fraction of the QS winds in this period is 17\%,
    and tends to increase with decreasing activity, accounting for 31\% of the winds in 2006.
 During the Cycle 23--24 minimum (2007 and 2008),
    the contribution of CHs (ARs) is about 31\% (15\%),
    while the QS contribution is $\sim 41$\%.
}

\item
{
Overall, in each phase of solar activity, the winds from CHs tend to be the fastest
    and associated with the lowest \oql\ ratios.
While both lower than CH winds, the speeds of AR and QS winds do not show a substantial difference.
A slightly more pronounced difference between AR and QS winds is seen in their \oql\ values, with AR winds
    tending to be slightly hotter.
As for the dependence on solar activity of the winds from the same sources,
       overall with decreasing activity the winds tend to
       have lower \oql\ ratios.
       }

\item The fractions occupied by the slow solar winds from different groups show a dependence on solar activity
    similar to the case where solar winds from all speed ranges are considered.
    This can also be said for the activity dependence of \oql\ values.
{During the minimum phase, the QS contribution to the slow wind
    is even more important than its overall contribution, amounting to $\sim 47$\%.}
\end{enumerate}


Our results suggest that the quiet Sun {is} an important source of the {ACE}
    solar winds around the cycle 23-24 minimum.
{
A further study dedicated to the examination of the properties of the source region and in situ properties
    of this particular QS wind is needed.}
Regarding the source regions, such properties as the magnetic field strength as well as magnetic topology will be of interest.
Regarding the in situ properties, the abundances of low first-ionization-potential (FIP) elements relative to their photospheric values
    will be informative~\citep{2005JGRA..110.7109F, 2009ApJ...691..760W}.
In addition, it will be worthwhile looking for direct signatures of outflow in the QS
    by examining the Doppler shifts with the emission lines
    measured with either SOHO/SUMER~\citep[e.g.,][]{2003A&A...399L...5X}
    or Hinode/EIS~\citep[e.g.,][]{2014ApJ...794..109F,2015NatCo...6E5947B}.

\begin{acks}
We would like to thank the anonymous referee for helpful
    comments.
We thank the ACE SWICS, SWEPAM, and MAG instrument
    teams and the ACE Science Center for providing the ACE
    data.
SOHO is a project of international cooperation
    between ESA and NASA.
Wilcox Solar Observatory data used in this study were obtained via the web site
    \url{http://wso.stanford.edu} courtesy of J.T. Hoeksema.
    The Wilcox Solar Observatory is currently supported by NASA.
This research is supported by the China 973
    program 2012CB825601, the National Natural Science Foundation of
    China (41174154, 41274176, and 41274178), and the Ministry of
    Education of China (20110131110058 and NCET-11-0305). BL is also
    grateful to the International Space Science Institute (ISSI) for
    providing the financial support to the international team on the
    origins of the slow solar wind.
\end{acks}

\bibliographystyle{spr-mp-sola}
\bibliography{ver4}


\clearpage

\begin{table}
\caption{Number of daily solar wind samples analyzed in each year.
} \label{tbl_polarity_all}
\begin{tabular}{clllll}     
  \hline                   
Year & All sources & CH winds & AR winds & QS winds & Undefined \\
  \hline
1999 & 237 (188) & 32 (27) & 148 (117) & 28 (23) & 29 (21)\\
2000 & 261 (221) & 50 (47) & 145 (125) & 36 (28) & 30 (21) \\
2001 & 272 (220) & 43 (39) & 155 (119) & 38 (30) & 36 (32)\\
2002 & 289 (259) & 65 (62) & 152 (135)  & 38 (35) & 34 (27)\\
2003 & 277 (259) & 142 (138) & 71 (65)  & 34 (32) & 30 (24)\\
2004 & 242 (211) & 63 (61) & 94 (81)  & 34 (30) & 51 (39)\\
2005 & 250 (201) & 78 (70) & 70 (61)  & 56 (34) & 46 (36)\\
2006 & 262 (200) & 60 (56)  & 55 (43)  & 94 (62) & 53 (39)\\
2007 & 257 (178) & 68 (57) & 36 (30)  & 114 (59) & 39 (32)\\
2008 & 259 (187) & 61 (58)   & 29 (27)  & 157 (92) & 12 (10)\\
  \hline
Sum  & 2606 (2124) & 662 (615) & 955 (803) & 629 (425) & 360 (281)\\
  \hline
\end{tabular}
\end{table}

\begin{table}
\caption{Number of daily slow solar wind samples analyzed in each
year.} \label{tbl_polarity_slow}
\begin{tabular}{clllll}     
  \hline                   
Year & All sources & CH winds & AR winds & QS winds & Undefined\\
  \hline
1999 & 188 (141) & 20 (15)   & 119 (89)  & 23 (18)  & 26 (19)\\
2000 & 204 (167) & 26 (23)   & 123 (103) & 29 (23)  & 26 (18)\\
2001 & 245 (194) & 37 (34)   & 139 (103) & 37 (29)  & 32 (28)\\
2002 & 238 (208) & 40 (37)   & 135 (118) & 32 (29)  & 31 (24)\\
2003 & 111 (96)  & 35 (31)   & 36 (32)   & 20 (18)  & 20 (15)\\
2004 & 194 (163) & 36 (34)   & 83 (70)   & 28 (24)  & 47 (35)\\
2005 & 172 (132) & 34 (29)   & 52 (44)   & 47 (28)  & 39 (31)\\
2006 & 205 (146) & 38 (34)   & 40 (28)   & 81 (50)  & 46 (34)\\
2007 & 187 (116) & 38 (28)   & 30 (25)   & 91 (42)  & 28 (21)\\
2008 & 169 (102) & 25 (22)   & 15 (13)   & 122 (61) & 7  (6) \\
  \hline
Sum  & 1913 (1465) & 329 (287) & 772 (625) & 510 (322) & 302 (231)\\
  \hline
\end{tabular}
\end{table}

\clearpage

   \clearpage

   \clearpage

\clearpage

\clearpage

   \clearpage

\clearpage

\end{article}

\end{document}